\documentclass[journal]{IEEEtran}
\usepackage[T1]{fontenc}

\usepackage{amsmath,amsfonts}
\usepackage{array}
\usepackage[caption=false,font=normalsize,labelfont=sf,textfont=sf]{subfig}
\usepackage{textcomp}
\usepackage{stfloats}
\usepackage{booktabs}
\usepackage{url}
\usepackage{graphicx}
\usepackage{pgfplots}
\usepackage{cite}
\usepackage{wrapfig}
\makeatletter
\def\@IEEEfigurecaptionsepspace{\vskip 0pt\relax}
\makeatother

\usepackage[nohyperlinks]{acronym}
\pgfplotsset{compat=1.18}
\definecolor{drawioBlue}{HTML}{6C8EBF}
\definecolor{drawioOrange}{HTML}{D79B00}
\definecolor{drawioGreen}{HTML}{82B366}
\definecolor{drawioRed}{HTML}{B85450}
\definecolor{drawioPurple}{HTML}{9673A6}
\definecolor{drawioGray}{HTML}{D6D6D6}

\acrodef{SP}{Strict Priority}
\acrodef{HP}{high-priority}
\acrodef{LP}{low-priority}
\acrodef{TXR}{transmit ring}
\acrodef{DSCP}{Differentiated Services Code Point}
\acrodef{DiffServ}{Differentiated Services}
\acrodef{DNC}{deterministic network calculus}
\acrodef{SNC}{stochastic network calculus}
\acrodef{PCP}{Priority Code Point}
\acrodef{TXQ}{transmit queue}
\acrodef{PDF}{probability density function}
\acrodef{MTU}{Maximum Transmission Unit}
\acrodef{TSN}{Time-Sensitive Networking}

\begin{document}

\title{Strict-Priority Packet Delay in Switches with Transmit-Ring Buffering}

\author{Yash Deshpande, Quirin Vogel, Wolfgang Kellerer.

%\thanks{Manuscript received Month Day, Year; revised Month Day, Year.}%
\thanks{Yash Deshpande and Wolfgang Kellerer are with the Chair of Communication Networks at the Technical University of Munich. Quirin Vogel is with the Department of Statistics, University of Klagenfurt, Austria.}}

% \markboth{IEEE Networking Letters}%
% {Author \MakeLowercase{\textit{et al.}}: Paper Title for IEEE Networking Letters}

\maketitle

\begin{abstract}
\ac{SP} scheduling is widely used at switch egress to provide low latency service to \ac{HP} traffic. Existing deterministic and stochastic latency models typically account for scheduler behavior and packet transmission, but omit a common switch implementation detail: the \ac{TXR} between the scheduler and the physical port. Because the switch must prepare the next packet before the current transmission completes, packets already placed in the \ac{TXR} can further delay \ac{HP} packets. This changes both the worst-case delay and the per-hop delay distribution of the \ac{HP} packets. This paper identifies this modeling gap, extends standard \ac{SP} latency models to include the \ac{TXR}, and validates the revised model through measurements on multiple switches. This paper also provides a measurement method for estimating the \ac{TXR} size, a parameter that is often not reported in switch datasheets. The resulting model provides a closer representation of switch behavior for systems that use \ac{SP} scheduling and require either delay bounds or delay distributions.
\end{abstract}
\acresetall


\section{Introduction}
Packet scheduling at switch egress is a standard mechanism for providing different quality-of-service classes to different flows. In industrial networks, safety-critical systems, and deterministic networking deployments, these scheduling mechanisms are often used to support latency guarantees for \ac{HP} traffic. Analytical methods such as deterministic network calculus model the scheduler and derive per-hop delay bounds from traffic and service assumptions~\cite{leboudec2001networkCalculus, jiang2008stochasticNetworkCalculus}.

\ac{SP} scheduling is a common choice for low-latency \ac{HP} traffic. It is available in many commercial off-the-shelf switches, including devices used in industrial, enterprise and data-center networks across. This availability has made \ac{SP} scheduling a frequent building block in deterministic networking studies, including joint scheduling and routing formulations on commodity switches~\cite{chameleonAmaury,diederich2025lcdn, grigorjewSP}. In these models, \ac{HP} traffic is assumed to receive service before \ac{LP} traffic whenever both are waiting at the scheduler.

Beyond worst-case bounds, stochastic network calculus and measurement-based analyses use per-hop delay distributions to estimate the probability of deadline violations~\cite{jiang2008stochasticNetworkCalculus}. Delay distributions are also useful for other timing-sensitive functions. For example, they can affect estimates of clock-synchronization error when packet delay variation contributes to delay assymetry~\cite{chaloupka_distribution}. Accurate per-hop delay models are therefore needed not only for strict guarantees, but also for probabilistic and statistical performance analysis.

\begin{figure}[!t]
\centering
\includegraphics[width=0.85\columnwidth]{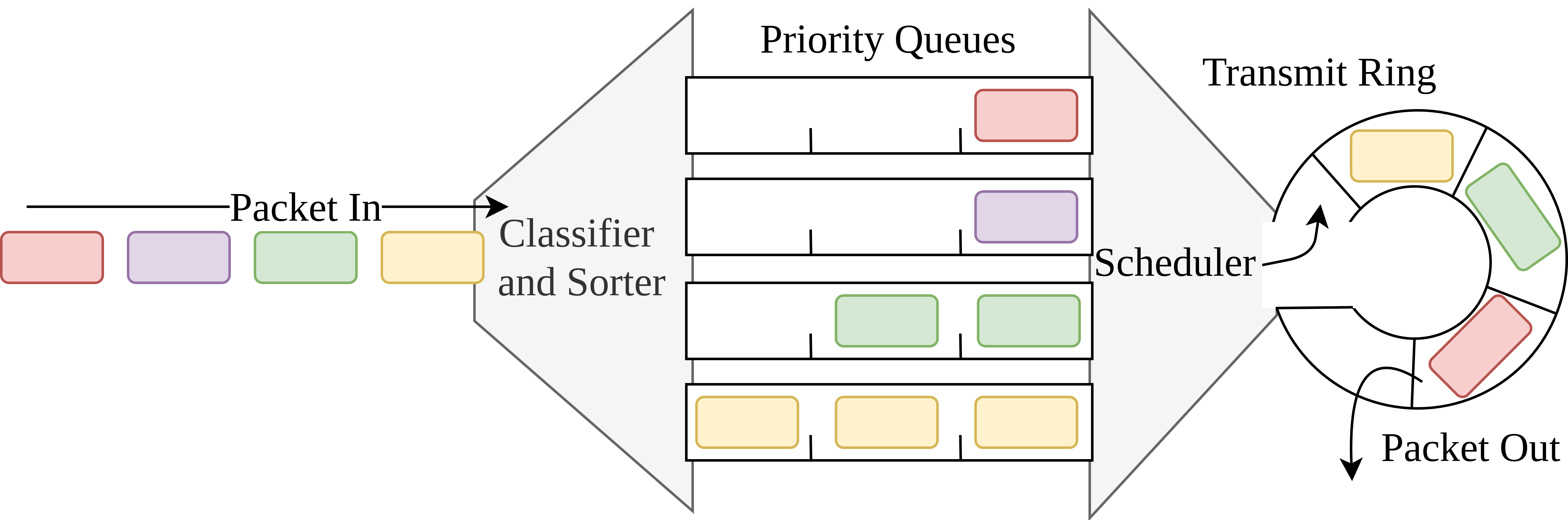}
\vspace{-0.2cm}
\caption{Egress architecture showing the post-scheduler \ac{TXR}.}
\label{fig:tx_queue_architecture}
\end{figure}

This paper shows that standard \ac{SP} latency models miss an implementation constraint that is present in most switches. To avoid idle time and packet-selection overhead at line rate, a switch may load one or more packets into a \ac{TXR} after the scheduler and before the physical egress port~\cite{cisco_srnd_network_infrastructure_2015}.
This architecture is illustrated in Fig.~\ref{fig:tx_queue_architecture}. 
%similar transmit descriptor rings and hardware queues are common in Ethernet controller designs~\cite{intelE810DatasheetTxDescriptors,linuxKernelRingBuffer}. 
Once a packet is placed in this \ac{TXR}, it can no longer be preempted by a later \ac{HP} arrival. As a result, an \ac{HP} packet can be delayed by \ac{LP} packets that have already passed the scheduler. This effect is not captured by models that treat the scheduler as the last queueing point before transmission.

The \ac{TXR} changes both deterministic and stochastic latency analysis. In the worst case, the \ac{HP} packet may wait behind multiple \ac{LP} packets already present in the \ac{TXR}. In the stochastic case, the delay distribution depends on the occupancy of the \ac{TXR} and on the interfering \ac{LP} traffic. The relevant \ac{TXR} size is usually not exposed as an operator setting, and it is often absent from switch datasheets. Reducing the ring size, when possible, may also increase CPU load or reduce forwarding efficiency.

This work makes two contributions: (1) We identify the \ac{TXR} as a source of discrepancy between standard \ac{SP} latency models and observed \ac{HP} packet delay. Measurements across different network switches show that this discrepancy exists in almost all the devices. (2) We extend the standard per-hop \ac{SP} delay model to include the \ac{TXR} size, for both worst-case latency and delay distributions. We validate the model using measurements and provide a measurement method for estimating the \ac{TXR} size.
The revised model can be used in latency analyses that require packet delay bounds or delay distributions for \ac{SP}-based networks.

\section{Background}
\label{sec:background}

Ethernet switches classify packets before enqueueing them at an egress port. The classification can use link-layer markings, such as the \ac{PCP} field in IEEE 802.1Q VLAN tags, or network-layer markings, such as the \ac{DSCP} field used by \ac{DiffServ}. The selected marking is mapped to one of the egress queues configured on the port. Commercial off-the-shelf switches commonly expose four or eight such priority queues per egress port. Other common egress schedulers include round robin, weighted round robin, and interleaved weighted round robin. Among these schedulers, \ac{SP} provides the lowest guaranteed latency to \ac{HP} packets.

Under \ac{SP} queueing, the scheduler always selects a packet from the highest non-empty priority queue. Therefore, an \ac{HP} packet is dequeued before any \ac{LP} packet that is still waiting at the scheduler. This rule is the basis for standard \ac{DNC} models of \ac{SP} scheduling, where \ac{LP} traffic can affect an \ac{HP} packet only through the packet already in transmission when the \ac{HP} packet arrives~\cite{leboudec2001networkCalculus}. \ac{SNC} models use related assumptions to derive a delay-violation probability or a delay distribution~\cite{jiang2008stochasticNetworkCalculus}.

\begin{figure}[!t]
\centering
\input{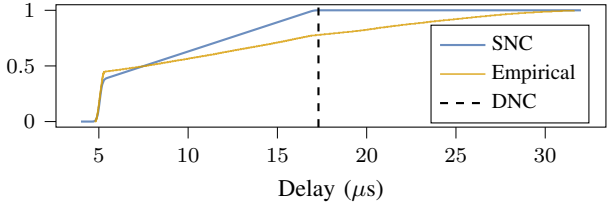}
\vspace{-0.2cm}
\caption{Empirical \ac{HP} latency CDF versus the CDF predicted by a standard \ac{SNC} model. The measured latencies show a different distirbution and violate the DNC bound. }
\label{fig:motivating_distribution}
\end{figure}

Fig.~\ref{fig:motivating_distribution} illustrates the discrepancy studied in this paper. In this example, three \ac{LP} flows send 1500-byte packets at 200~Mbps each on a 1~Gbps link, while one \ac{HP} probe packet arrives at a random time once per second. We measure 27,000 \ac{HP} packet latencies and compare the empirical distribution with the distribution predicted by standard network-calculus models. Under the standard \ac{DNC} model, the worst-case queueing delay is the non-preemptive blocking time of one packet, $d_{q}^{\mathrm{DNC}} = L/C = 1500\cdot8/10^9 = 12~\mu\mathrm{s}$, where $L$ is the packet size and $C$ is the link capacity.  
The corresponding total delay is the maximum processing delay, $d_{\text{proc}}^{\max}=5.32\mu\mathrm{s}$, added to this value.
The $d_{\text{proc}}^{\max}$ value is measured over 10,000 \ac{HP} probe samples without \ac{LP} flows. 
Alternatively, if available, one can refer the RFC2544~\cite{rfc2544} test results of the switch.   
The \ac{SNC} model predicts that the total-delay \ac{PDF} is a mixture distribution: with probability $0.4$ it is Gaussian, representing probes that are not queued, and with probability $0.6$ it is the sum of a Gaussian term and a uniform random variable on $[0,d_{q}^{\mathrm{DNC}}]$.
The two distributions differ in the queuing regime.

The same discrepancy affects worst-case analysis. If the traditional \ac{DNC} model is used as the target latency, the predicted bound can be lower than the measured latency for $22\%$ of the probes in this experiment. This means that the model can underestimate the latency target even when the scheduler is configured as expected.
The DNC bounds are supposed to be \textit{strict}, and therefore a violation to it untenable for certain applications. 

The same effect was observed in prior work, but it was treated as a priority-queue overhead~\cite{amauryEmpirical}. 
That study measured line-rate saturated \ac{LP} queues and did not analyze the full latency distribution. 
It also did not identify why the overhead differs across switches.

The cause is the \ac{TXR} located after the scheduler and before the physical egress port as shown in Fig.~\ref{fig:tx_queue_architecture}.  
Before the currently transmitted packet completes, the scheduler can already dequeue the next packet from the priority queues and place it in the \ac{TXR}. This allows the switch to prepare the next transmission in advance and avoids egress-port idle time. 
%This reduces packet-selection work on the critical path.  
Once an \ac{LP} packet has entered this queue, a later \ac{HP} packet cannot overtake it, even though \ac{SP} scheduling would have selected the \ac{HP} packet if both packets were still in the priority queues.

The \ac{TXR} capacity is usually a number of packets, and this value determines how many already-dequeued packets can block a later \ac{HP} packet. 
Over a variety of switches measured in Sec.~\ref{subsec:txr_size}, we determined that the TXR size is between 2-4 packets. 
Only for 1G \ac{TSN} switches capable of IEEE 802.1Qbu frame preemption, TXR size was 1 packet. 
In the next section, we formulate the problem, introduce the notation, and give the main results for the packet-delay bound and delay distribution as functions of the \ac{LP} traffic characteristics. The following section describes the measurement setup and validates the proposed model. To the best of our knowledge, this is the first work to provide both a worst-case and a probabilistic queue-delay model for \ac{SP} scheduling that explicitly includes the post-scheduler \ac{TXR}.

\section{System Model}
\label{sec:system_model}

We consider a single egress port configured with \ac{SP} scheduling. From the perspective of \ac{HP} packets, all queues with lower priority than the \ac{HP} queue can be abstracted into a single aggregate \ac{LP} queue~\cite{leboudec2001networkCalculus, amaurynetworkcaclus}. This abstraction is sufficient because an \ac{HP} packet is affected only by lower-priority packets that are either already in service or already placed in the post-scheduler \ac{TXR}.

We assume that at most one \ac{HP} packet is present in the system at any time. The \ac{HP} packet arrives at a random time and serves as a probe for the instantaneous delay caused by the \ac{LP} traffic and the \ac{TXR}. If this assumption is relaxed, an arrival process for the \ac{HP} flow(s) must additionally be specified, since \ac{HP} packets may then queue behind other \ac{HP} packets.

Let $C$ denote the link capacity. We consider $N$ enumerated \ac{LP} flows indexed by $f\in{1,\ldots,N}$. Each \ac{LP} flow $f$ is characterized by a fixed packet size $L_f$ in bytes and a rate $\rho_f$. The rate $\rho_f$ is expressed as a fraction of the link capacity $C$ and therefore satisfies $0\leq \rho_f \leq 1$. 
The \ac{TXR} has a capacity of $B$ packets. We assume that the priority queues before the scheduler have infinite capacity. Buffer overflow is excluded by requiring that the aggregate \ac{LP} load $\rho$ satisfies, $\rho=\sum_{f=1}^{N} \rho_f \leq 1$.
This condition ensures that the aggregate \ac{LP} input rate is sustainable by the egress link in the long term.
\section{Analysis}
\label{sec:analysis}

\subsection{Without TXR}

\subsubsection{Stochastic}

Let $D_{\mathrm{HP}}$ be a random variable that denotes the delay experienced by an \ac{HP} packet. We model this delay as the sum of two independent components, $D_{\mathrm{HP}} = D_0 + D_{\mathrm{LP}}$, where $D_0$ represents the baseline system delay and $D_{\mathrm{LP}}$ represents the additional delay caused by LP traffic already occupying the server or the \ac{TXR}.

The baseline delay $D_0$ captures processing, propagation, and other stochastic system effects unrelated to queue occupancy. It is modeled as a Gaussian random variable $D_0 \sim \mathcal{N}(\mu,\sigma^2)$, with mean $\mu$ and variance $\sigma^2$. 

The additional delay component $D_{\mathrm{LP}}$ is modeled as a mixed random variable determined by the instantaneous state of the LP traffic at the arrival instant of the \ac{HP} packet. Under \ac{SP} scheduling with TXR, an arriving \ac{HP} packet can only be blocked by a LP packet that is already in transmission. Due to the random arrival instant of the \ac{HP} packet, the remaining service time of a blocking packet is uniformly distributed over the serialization interval of that packet.
For a LP flow $f$, the serialization delay is $T_f = \frac{L_{f}\cdot 8}{C}$.
The blocking delay caused by flow $f$ is therefore modeled as a uniformly distributed random variable from 0 to the serialization dely, $U_f \sim \mathcal{U}(0,T_f)$, 
because the HP packet can arrive at anytime during the transmission of the LP packet. 
With probability $1-\rho$, no LP packet is encountered and therefore no additional blocking delay occurs. With probability $\rho_f$, the \ac{HP} packet is blocked by a packet belonging to flow $f$.

Consequently, the random variable $D_{\mathrm{LP}}$ follows the mixed distribution
\[
D_{\mathrm{LP}} \sim
(1-\rho)\,\delta(x)
+
\sum_{f=1}^{N}\rho_f\,\mathcal{U}(0,T_f),
\]
where $\delta(x)$ denotes the Dirac delta distribution representing zero blocking delay.

The overall delay distribution of the \ac{HP} packet is therefore obtained as the convolution of the Gaussian baseline delay with the mixed blocking-delay distribution,
\[
f_{D_{\mathrm{HP}}}(x)
=
(1-\rho)f_G(x)
+
\sum_{f=1}^{N}\rho_f
\left(
f_G * f_{U_f}
\right)(x),
\]
where $f_G(x)$ denotes the Gaussian probability density function and $*$ denotes convolution.

The convolution can be explicitly solved and simplified as
\begin{equation}\label{eq:one-schedule}
        f_{D_{\mathrm{HP}}}(x)
=
(1-\rho)f_G(x)
+
\sum_{f=1}^{N}\rho_f\frac{F_G(x)-F_G(x-T_f)}{T_f}\, ,
\end{equation}
where $F_G$ is the associated Gaussian cummulative distribtuion function.

This model captures the key behavior of \ac{SP} scheduling without a post-scheduler \ac{TXR}: in the absence of LP blocking, the delay follows the baseline Gaussian behavior, whereas the presence of a LP packet introduces an additional uniformly distributed serialization delay whose magnitude depends on the corresponding packet size.

\subsubsection{Deterministic}
\ac{DNC} provides a deterministic upper bound on the \ac{HP} packet delay. In this case, the stochastic baseline delay component is replaced by its maximum observed value %\footnote{Strictly speaking, a Gaussian random variable does not possess a finite upper bound. However, in practical systems, the Gaussian component is typically obtained as a statistical fit to measured delay samples. Therefore, a finite maximum delay can be defined from the collected data set and used as a deterministic approximation.},
denoted by $d_{\text{proc}}^{\max}$.%
One can also use a percentile based cutoff on a gaussian with $\mu,\sigma^2$ to obtain $d_{\text{proc}}^{\max}$. 

Under \ac{SP} scheduling, the worst-case blocking occurs when the arriving \ac{HP} packet encounters the largest possible LP packet just about to start a transmission. Let $F \in \arg\max_{f\in\{1,\ldots,N\}} L_f$ denote an \ac{LP} flow with maximum packet size. The resulting deterministic upper bound on the \ac{HP} packet delay is therefore given by
\begin{equation}\label{eq:dnc_old}
d_{\mathrm{HP}}^{\mathrm{DNC}}
=
d_{\text{proc}}^{\max}
+
T_{F}.        
\end{equation}

This bound corresponds to the classical non-preemptive SP worst-case analysis, where the \ac{HP} packet may be blocked by at most one LP packet from flow $F$.

\subsection{With TXR}

\subsubsection{Stochastic}
In the previous model, only a single LP packet could block the arriving \ac{HP} packet. However, when a \ac{TXR} of capacity $B$ packets is present, multiple LP packets may already occupy the transmission pipeline ahead of the arriving \ac{HP} packet.

We now define the stochastic \ac{TXR} model by explicitly accounting for the occupancy contribution of each LP flow.

Let $K \in \{0,1,\ldots,B\}$ denote the random number of LP packets located ahead of the arriving \ac{HP} packet, including packets currently in transmission and packets buffered inside the \ac{TXR}. 
%The state $K=B$ represents the aggregated event that the number of blocking packets is greater than or equal to the \ac{TXR} capacity, i.e.,
%\[
%\Pr(K=B) = \Pr(K \geq B).
%\]
The occupancy distribution of $K$ depends on the interaction between the LP flows, including their arrival characteristics and batching (burst) behavior. 
In this work, we do not explicitly model these arrival processes. Instead, the aggregate occupancy behavior is represented by the probability mass function $p_k = \Pr(K=k)$
with $\sum_{k=1}^{B} p_k = \rho$, i.e th total steady state probability that a packet is blocked is the aggregate rate of all the flows. 

Conditioned on the event $K=k$, there exist multiple possible compositions of the $k$ blocking packets across the $N$ LP flows.
We need to be careful and distinguish the serialization delay of a packet being served and the packets already loaded. Denote the flow former by $g$. Let furthermore $\mathbf{n}^{(k)} = (n_1,\ldots,n_N)$ denote one such composition of loaded packets, where $\sum_{f=1}^{N} n_f = k-1$, and $n_f$ represents the number of packets belonging to flow $f$.
Each valid composition is associated with conditional probabilities, $\chi_f=\Pr\!\left(g=f\mid K\ge 1\right),$
as well as $\pi_{\mathbf{n}}^{(k,f)}=\Pr\!\left(\mathbf{n}^{(k)} \mid K=k, g=f\right),$
satisfying
\[
\sum_{f=1}^N\chi_f=1\quad\textnormal{and}\quad\sum_{\mathbf{n}^{(k)}} \pi_{\mathbf{n}}^{(k,f)} = 1,
\]

where the summation is taken over all valid compositions of $k$ packets across the $N$ flows. 

% Note that we do not necessarily assume $\chi$ and $\pi$ to be related. If $L_1=1500$ and $L_2=64$ and $\rho_1=\rho_2$, we would expect $\chi_1=\chi_2=1/2$ since both type of packets occupy the flow equal amount of time. Indeed, if the system is in steady state, it would be reasonable to put

% \[
% \chi_f=\frac{\rho_f}{\rho}\, .
% \]
% However, since packets sizes might not be equal, the above relation might not hold for $\pi$. Indeed, if $\rho_f$ is the flow rate, then this implies an arrival rate of $\lambda_f=\rho_f/T_f=\rho_f C/(8L_f)$. In our example, this would give $\lambda_1\approx 4.1\%$ and $\lambda_2\approx 95.9\%$. If I understand you correctly Yash, it is package based, so we would expect a bias towards packets of type $64$ bytes. Hence, we would expect (under independence between flows and stationarity)
% \[
% \pi_{\mathbf{n}}^{(k,f)}\sim \mathrm{Mult}\left(\frac{\lambda_1}{\sum_f \lambda_f},\ldots,\frac{\lambda_N}{\sum_f \lambda_f}\right)\, .
% \]
% However, in this paper, we keep the formalism abstract, to allow the modeling for different situations (independence, burst periods, etc..).

Conditioned on a specific composition $\mathbf{n}^{(k)}$, the total LP blocking delay is therefore given by
\begin{equation}\label{eq:unifTrans}
D_{\mathrm{LP}}^{(k,\mathbf{n})}
=\mathcal{U}(0,T_g)+
\sum_{f=1}^{N}
\sum_{i=1}^{n_f}
T_{f,i}=\mathcal{U}(0,T_g)+
\sum_{f=1}^{N}n_fT_f\, .
\end{equation}

Another phenomenon that is not modelled in typical DNC and SNC so far is the arbitration latency (AL) ~\cite{arbitration_overhead}. 
Modern switches often implement priority queues using shared internal buffers or virtual queues. When the \ac{SP} scheduler selects one packet from these queues, the selection step can introduce an overhead in the form of AL. 
We also measured this overhead in our tests and the following phenmonena was observed: instead of continous translated uniform $k$-dependent blocks (see \eqref{eq:unifTrans}), a gap between the uniform components of the law of $D_{\mathrm{HP}}$ appear. Mathematically, this can by expressed by introducing a parameter $\delta_{\mathrm{AL}}>0$ into the model. 
The unconditional LP blocking-delay distribution is then obtained as a weighted mixture over all occupancy states and all valid flow compositions as well as incorporating the AL. 
The PDF is given by, 
\begin{multline}
f_{D_{\mathrm{HP}}}(x)
=(1-\rho)f_G(x)+\sum_{k=1}^{B}
p_k\sum_{f=1}^N
 \chi_f\sum_{\mathbf{n}^{(k)}}
\pi_{\mathbf{n}}^{(k,f)}\\
\times
\frac{F_G(x-S_{\mathbf{n}^{(k)},\mathrm{AL}})-F_G(x-S_{\mathbf{n}^{(k)},\mathrm{AL}}-T_f)}{T_f}\,
\label{eq:final_boss}
\end{multline}
with $S_{\mathbf{n}^{(k)},\mathrm{AL}}=S_{\mathbf{n}^{(k)}}+(k-1)\delta_{\mathrm{AL}}$, gives the additional translation of the density by $\delta_{\mathrm{AL}}$. 
This formulation extends the single-packet blocking model to finite-capacity post-scheduler buffering by incorporating the stochastic occupancy of the \ac{TXR}, the flow-dependent composition of the blocking packets and the AL. 
%Fitting of this more complicated model (i.e., estimating $\delta_{\mathrm{AL}}$) can be done as before. 

\subsubsection{Deterministic}

The worst-case delay is obtained when the packets ahead of the arriving \ac{HP} packet have the largest serialization time. With a post-scheduler \ac{TXR} of capacity $B$ packets, the \ac{HP} packet can be blocked by up to $B$ already-dequeued \ac{LP} packets in addition to the baseline processing delay.
Therefore, the deterministic upper bound with the \ac{TXR} is
\begin{equation}\label{eq:dnc_updated} 
d_{\mathrm{HP}}^{\mathrm{DNC,TXR}}
=
d_{\text{proc}}^{\max}
+
B T_{F}
+ (B-1)\delta_{\mathrm{AL}}
.
\end{equation}

This bound extends the classical non-preemptive \ac{SP} bound by replacing the single blocking packet with the largest possible set of $B$ packets already admitted into the \ac{TXR} along with the AL for $B-1$ gaps.
Both updated models, Eq~\ref{eq:final_boss} and ~\ref{eq:dnc_updated} can be reduced to their non-TXR form (Eq.~\ref{eq:one-schedule}, \ref{eq:dnc_old}) by plugging $B=1$ and $\delta_{\mathrm{AL}}=0$.   
\section{Measurement and Evaluation}
\label{sec:measurement_and_evaluation}

We use the single-NIC loopback method to measure the total latency at the switch witout the need for any clock synchronization~\cite{yash_infocom}. 
Such a method uses the hardware timestamping mechanism in the NIC and reduces the error to $\pm8ns$. 
For the 1G measurments we use an intel i350 and for the 10G measurements we use an intel x520 NIC. 
The LP flows can be generated by upto 4 seperate flows coming from seperate ports using \textit{iperf}. 
The non-blocking parameters for the mean and variance of $D_0$ are first collected over 10,000 samples without any LP traffic. 
Each real measurement is taken from at least 27,000 samples.

\subsection{TXR Size} 
\label{subsec:txr_size}
\begin{figure}[!t]
\centering
\input{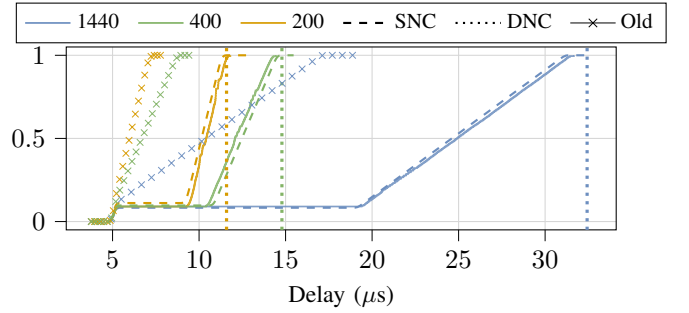}
\vspace{-0.6cm}
\caption{The TXR capacity is in packets, not bytes; Therefore, smaller packets experience a smaller overall delay. The models proposed in Eq.~(4) and (5) shown in dashed lines fit the observed data in solid lines in contrast to the old model Eq.~(1) in cross marks. }
\label{fig:txr_size}
\end{figure}

In this test all the LP flows transmit UDP packets such that their aggregate rate equals the link capacity of 1 Gbps. 
We then vary the packet size from 200 to 1500 bytes to ascertain that the TXR is indeed in terms of packets and not bytes, which is the case for priority queues. 
The results are shown in Fig.~\ref{fig:txr_size}. 
If the TXR were defined in bytes, it would remain full in all three tests and the probe packet would be queued behind approximately the same number of bytes. 
This would produce nearly equal delays for all packet sizes, which is not observed. 
Instead, the delay decreases for smaller packets, showing that the probe packet is blocked by a fixed number of packets rather than a fixed number of bytes. 
Using the $T_f$ values for each packet size, the observed separation between the curves gives a TXR size of $B=2$ packets and $\delta_{\mathrm{AL}}=2.1\mu$s. 
If the TXR size and AL is unknown, such a test can be applied to determine its value. 
Substituting these test conditions into Eq.~(4), the aggregate load is $\rho\approx1$ and $B=2$.  
Because the LP flows have identical packet sizes and rates within each experiment, $\mathbf{n}^{(k)}$ has a single realization and $p_{B}\approx1$. 
With these values, the model closely matches the measured CDFs for all three packet sizes as seen in Fig~\ref{fig:txr_size}, comparing the dashed curves with the observed data in solid curves for each packet size.
The vertical lines are the DNC bounds from Eq.~\ref{eq:dnc_updated} which also closely match the obtained results. 
In contrast, the old model (Eq.~(1)) in cross marks is noticeably divergent from the observed values. 

The measurements presented above were obtained using an FS-S2805S 1G switch and show that the \ac{TXR} capacity is packet-based rather than byte-based, with a default size of two packets for this switch.

\begin{wraptable}{r}{0.48\columnwidth}
\vspace{-0.8\baselineskip}
\centering
\caption{Default \ac{TXR} size and AL}
\label{tab:txr_size_placeholder}
\scriptsize
\setlength{\tabcolsep}{1pt}
\renewcommand{\arraystretch}{0.65}
\begin{tabular}{@{}lccc@{}}
\toprule
Switch & Rate & TXR & AL \\
       & (Gb/s) & Size & ($\mu$s) \\
\midrule
FS S2805S & 1  & 2 & 2.1 \\
Dell S4048    & 10 & 3 & 1.7 \\
FS S5850    & 10 & 3 & 1.3 \\
Edgecore AS7726-32X     & 25 & 4 & 1.1 \\
\bottomrule
\end{tabular}
\end{wraptable}

We also conducted additional measurements across multiple 1G and 10G switches. 
The default \ac{TXR} sizes and AL overhead observed in those experiments are summarized in Table~\ref{tab:txr_size_placeholder}. 
Overall, the 10G and faster switches exhibit larger \ac{TXR} sizes, which increases the potential blocking delay experienced by \ac{HP} traffic under strict priority scheduling. 
The switches for which this discrepancy was not present were the Kontron D10 \ac{TSN} switch and the Relyum \ac{TSN} Bridge switch. 
Both devices support IEEE 802.1Qbu frame preemption, which requires a different egress architecture to stop an ongoing LP frame.
However, for Kontron D10, we still measured an AL of 1.4$\mu$s. Thus the updated model with $B=1$ is still useful for such cases. 

\subsection{TCP Approximation}

Estimating the exact values of $p_k$, $\chi_f$, and $\pi_{\mathbf{n}}^{(k,f)}$ is difficult under realistic traffic conditions. 
These quantities depend on complex inter-flow dynamics, flow burstiness, and packet inter-arrival times, which are not captured by the average flow rates alone. 
However, a useful approximation can be made when all \ac{LP} flows are TCP flows with known rates. 
TCP traffic is inherently bursty because congestion-window transmission often releases multiple packets back-to-back after an acknowledgement or after the sender receives a window update, creating short packet trains rather than isolated packet arrivals. 
Moreover, for bulk TCP transfers, the packet size is typically close to the \ac{MTU}. 
Thus, the serialization delay of each \ac{LP} packet can be approximated by the \ac{MTU} serialization delay, while the relative contribution of each flow can be inferred from its known rate.

\begin{figure}[!t]
\centering
\input{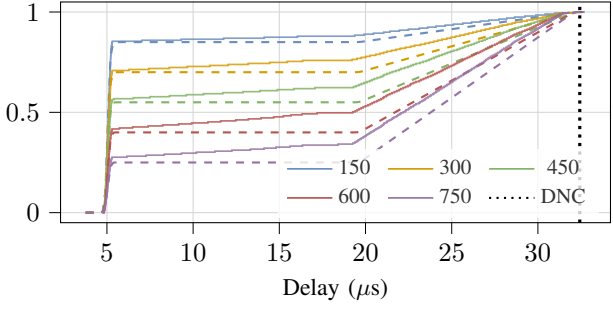}
\vspace{-0.2cm}
\caption{Comparison for TCP approximated model (dashed) and the corresponding measurements (solid). The approximation performs better as the aggregate TCP rate from the LP flows decreases. }
\label{fig:tcp_approximation}
\end{figure}

Fig.~\ref{fig:tcp_approximation} compares this approximation with the measured CDFs for different aggregate TCP rates in Mbps on a link with 1Gbps capacity. 
Somewhat counterintuitively, the approximation improves as the aggregate \ac{LP} rate decreases as the relative burst of the smaller flows is high. 
%At high load, the TXR is more likely to be filled by TCP bursts, so the dominant blocking event is $K=B$ and the approximation $p_B\approx1$ captures the measured behavior well. 
%At lower load, the port is less busy and the TXR is not always full. 
The approximation places the region before the linear part of the CDF as flat, which corresponds to setting $p_1=0$. 
However, even with bursty TCP arrivals, there remains a small probability $p_1$ that the probe packet is blocked by only one \ac{LP} packet, which explains the discrepancy between the model and the empirical measurements.
However, SNC provides delay bounds and the figure shows that the anayltical approximation does not violate the measured values, i.e., the dashed lines always remain below the solid ones.  
A posterior fit on the empirical data shows that $p_1=0.1\rho$ and $p_2=0.9\rho$ provides tigher bounds. 
%This produces additional probability mass before the main linear region and explains the larger gap between the approximation and the observed data at lower rates.

\begin{figure}[!t]
\centering
\input{figures/eight_dataset_cdf_comparison.tex}
\vspace{-0.2cm}
\caption{Effect of UDP burstiness on the measured delay for two packet sizes.}
\label{fig:burstiness_comparison}
\end{figure}

\subsection{UDP with Burst}

For UDP tests, \textit{iperf} usually spaces packets approximately evenly over time. Its burst parameter can instead select how many packets are sent back-to-back before the sender waits for the next inter-arrival interval, while preserving the configured average flow rate. We perform four such tests for each of two packet sizes, 700 and 1500 bytes, using burst values of 1, 2, 10, and 100 packets. 
The LP flows were started at random times to allow for some interference between the batches of each flow. 
Fig.~\ref{fig:burstiness_comparison} shows that, when burstiness is low, the resulting delay distribution can be approximated by a linear model i.e., $p_1=p_2=\frac{\rho}{2}$. 
As burstiness increases, however, the probability that the \ac{TXR} is filled increases, and the dominant blocking event shifts toward $p_B$. In practice, it is difficult to set up a measurement in which the stationary values of $p_K$ can be directly controlled and independently verified. We leave further investigation of these effects and their parameter estimation for future work.

\section{Conclusion}

This paper identified a \ac{TXR}-related discrepancy in packet-delay modeling for Ethernet switches that perform \ac{SP} scheduling, where \ac{LP} packets that have already been admitted into the post-scheduler transmit ring can continue to block an arriving \ac{HP} packet. 
We provided both stochastic and deterministic models for this effect and validated them through \ac{SNC} and \ac{DNC} comparisons.

By incorporating the post-scheduler \ac{TXR} into both deterministic and stochastic analyses, the proposed models make \ac{DNC}- and \ac{SNC}-based evaluations more representative of deployed network switches. This brings analytical results closer to measured system behavior and can improve the reliability of latency guarantees and violation-probability estimates for systems that rely on strict-priority scheduling.

\bibliographystyle{IEEEtran}
\bibliography{references}

\end{document}